\documentclass{moriond}

\def\Journal#1#2#3#4{{#1} {\bf #2}, #3 (#4)}

\def\NPB{{\em Nucl. Phys.} B}
\def\PLB{{\em Phys. Lett.}  B}

\def\PRD{{\em Phys. Rev.} D}

\def\PR{{\em Phys. Rept.}}
\def\AHEP{{\em Adv. High Energy Phys.}}
\def\JHEP{{\em JHEP}}
\def\tp{{\cal O}_{\tt CP}}
\def\gsim{\lower0.5ex\hbox{$\:\buildrel >\over\sim\:$}}
\def\lsim{\lower0.5ex\hbox{$\:\buildrel <\over\sim\:$}}

\def\be{\begin{equation}}
\def\ee{\end{equation}}
\def\bea{\begin{eqnarray}}
\def\eea{\end{eqnarray}}

\newcommand{\Photo}{\vspace{0.5cm} \em Talk given at the $57^{\rm th}$ Recontres de Moriond 2023, EW session (flavour physics) \vspace{0.2cm}\\ 
Based on: "Generic tests of CP-violation in high-$p_T$ multi-lepton signals at the LHC and beyond”, \\
Y. Afik, S. Bar-Shalom, K. Pal, A. Soni, J. Wudka, arXiv: 2212.09433 (hep-ph)}

\begin{document}
\vspace*{4cm}
\title{CP tests for high-$p_T$ multi-leptons}

\author{ Shaouly Bar-Shalom}

\address{Physics Department, Technion--Institute of Technology, Haifa, Israel}

\maketitle\abstracts{
We propose a generic and model-independent search strategy for probing tree-level CP-violation in inclusive multi-lepton signals and we extend the standard expression for tree-level CP-violation in scattering processes at the LHC to include the cases where the initial state in not self-conjugate. We then use TeV-scale 4-fermion operators of the form $tu\ell\ell$ and $tc \ell \ell$ with complex Wilson coefficients as an illustrative example and show that it may generate ${\cal O}(10\%)$  CP asymmetries that should be accessible at the LHC with an integrated luminosity of ${\cal O}(1000)$ fb$^{-1}$.}

\section{Introduction}

CP-violation (CPV) may be the key to a deeper understanding of particle physics and the evolution of the universe. It has far-reaching implications for cosmology since it 
is needed to explain the observed baryon asymmetry of the universe, while CPV within the SM is insufficient for that.
Furthermore, CP is not a symmetry of nature, so that 
on general grounds, one expects any generic new physics to entail Beyond the SM (BSM) CP-odd phase(s), e.g., 
SUSY models, Mutli-Higgs models, Leptoquarks, Vector-like fermions etc ...

In this talk, which is based on \cite{ourpaper}, we re-examine the formulation of tree-level CP-violating effects in scattering processes at the LHC and we 
introduce a new term to the "master" CPV expression, which takes into account "fake" CP-violating effects that arise when the initial state in not self-conjugate.
Based on that, we motivate an inclusive search for CPV in tri-lepton and four-leptons events at the LHC: 
\begin{eqnarray}
    && pp \to \ell^{\prime -} \ell^+ \ell^- + X_3 ~, \nonumber \\
    && pp \to \ell^{\prime +} \ell^- \ell^+ + \bar{X}_3 ~, \nonumber \\
    && pp \to \ell^{\prime +} \ell^{\prime -} \ell^+ \ell^- + X_4 \label{multileptons}     ~,
\end{eqnarray}
where $\ell,\ell^\prime=e,\mu,\tau$ (preferably $\ell \neq \ell^\prime$, see below) and $X_3$, $\bar{X}_3$ and $X_4$ contain in general jets and missing energy.
These include the $e^\pm \mu^+ \mu^-$ and $\mu^\pm e^+ e^-$ final states for $\ell,\ell^\prime=e,\mu$ and similarly for the pairs $\ell,\ell^\prime=e,\tau$ and $\ell,\ell^\prime=\mu,\tau$ 
and also the 3-flavor $e \mu \tau$ final state. 

Indeed, the multi-leptons signals of Eq.~\ref{multileptons} are excellent test ground for new physics (NP), 
being the "end-signals" of many interesting processes such as  $pp \to t \bar t V, ~t \bar t H, tV, tt \bar t \bar t, VV, VVV ...$ ($V=Z,W$).  
In particular, they are rich and clean signals in the hadronic environment of the LHC  
and are sensitive to many types of underlying NP, e.g.,  lepton-flavor violation, lepton universality violation,
lepton-number violation (same-sign leptons), CP violation etc ...
Moreover, High-$p_T$ (TeV energies) charged-leptons are still relatively unexplored and are relatively easily identifiable objects. 
All that and more were explored recently in detail \cite{ours1,ours2,ours3,ours4,ours5}. 

\section{CPV in multi-leptons events}\label{subsec:CPVML}

We construct below generic tests of CP in multi-lepton processes, 
focusing on the tri-lepton events of Eq.~\ref{multileptons}, but our derivation below applies also 
to the 4-leptons signals of Eq.~\ref{multileptons}. 

We would like to emphasize two important points regarding CPV test in multi-leptons events at the LHC:
\begin{itemize}
\item Sizeable, say ${\cal O}(1\%)$ manifestation of CPV in these multi-leptons events 
will be an unambiguous indication of NP, since the SM background to CP in these processes is expected 
to be negligible (un-observably small); the CPV effects from the SM's CP-odd phase of the CKM matrix
can arise only from EW processes at a higher loop order. 
\item CP studies at the LHC are more complicated, since the initial state is not self-conjugate. 
We thus developed a rigorous formula for testing CPV in an LHC-like environment, which includes 
a modification to the CP “master” formula.
\end{itemize}

\subsection{Constructing observables for tree-level CPV}\label{subsec:CPV1}

Consider the underlying hard process (parton level) for tri-lepton production at the LHC (and its CC channel):
\begin{equation}
a b \to \ell^{\prime -} \ell^+ \ell^- ~~, ~~ \bar{a} \bar{b} \to \ell^{\prime -} \ell^+ \ell^-  \label{eq:3lhard}
\end{equation}

For CPV, at least two interfering amplitudes are required ($M_1$ and $M_2$):
\begin{eqnarray}
    {\cal{M}}_{a b \to \ell^{\prime -} \ell^+ \ell^-} &=& M_1 e^{i \left(\phi_1 + \delta_1 \right) } + M_2 e^{i\left( \phi_2 + \delta_2 \right) }  \label{M} ~,
\end{eqnarray}
so that for the charged-conjugate channel we have
\begin{eqnarray}    
    \bar{\cal{M}}_{\bar{a} \bar{b} \to \ell^{\prime +} \ell^- \ell^+} &=& M_1 e^{i \left(-\phi_1 + \delta_1 \right) } + M_2 e^{i\left( - \phi_2 + \delta_2 \right) } \label{Mbar}~,
\end{eqnarray}
where $\phi_{1,2}$ and $\delta_{1,2}$ are CP-odd and CP-even phases, respectively. We will use below 
the relative phases $\Delta\phi = \phi_1 -\phi_2$, $\Delta \delta=\delta_1 - \delta_2$. 

It is useful to classify CP observables according to their transformation properties under "naive time reversal" ($T_N$), where 
$T_N$: time $\to$ -time.\cite{ourreview} This is presented in Table~\ref{tab:tab1}. In particular, a $T_N$-odd CP-odd observable 
requires only a CP-odd phase, which could arise already at the tree-level, while a  $T_N$-even CP-odd observable needs also a non-vanishing CP-even phase from 
final state interactions (FSI) which is typically of a higher order. 

\begin{table}[htb]
\caption{Classification of CP-odd observables under naive time reversal $T_N$, 
for the above tri-lepton processes, where $\Delta\delta \equiv \delta_1-\delta_2$, $\Delta\phi \equiv \phi_1-\phi_2$ and 
$\delta_i,\phi_i$ are defined within the amplitudes  ${\cal{M}}_{a b \to \ell^{\prime -} \ell^+ \ell^-} $ and 
$\bar{\cal{M}}_{\bar{a} \bar{b} \to \ell^{\prime +} \ell^- \ell^+} $ in Eqs.~\ref{M} and \ref{Mbar}.
\label{tab:tab1}}
\begin{center}
\begin{tabular}{c|c|c}
 Type & $T_N$-odd (CP-odd) & $T_N$-odd (CP-even)  \tabularnewline
\hline \hline
\
CP-asymmetry & $A_{CP} \propto \cos\Delta\delta \sin\Delta\phi$ & $A_{CP} \propto \sin\Delta\delta \sin\Delta\phi$ \tabularnewline
\hline 
\
Required phases & only CP-odd & Both CP-odd \& CP-even \tabularnewline
\hline 
\
Sensitivity & tree-level CPV & CP-even phase from FSI  \tabularnewline
\
 &  & (higher order effect)  \tabularnewline
 \hline 
\
Expected size & ${\cal O}(10\%)$ & ${\cal O}(0.1\%)$  \tabularnewline
\hline \hline
\end{tabular}
\end{center}
\end{table}

Therefore, in order to probe CPV in the tri-lepton channels of Eq.~\ref{eq:3lhard}, we have constructed in \cite{ourpaper} CP-odd observables 
based on the following $T_N$-odd 
triple products (TP):
\footnote{Useful TP correlations for CP studies in scattering processes at the LHC, which involve leptons with jets momenta, e.g., in $t \bar t$ and vector-bosons production, have been also suggested in~\cite{TP1,TP2,TP3,TP4,TP5,TP6}.}
\begin{eqnarray}
\tp = \vec{p}_{\ell^{\prime -}} \cdot \left( \vec{p}_{\ell^+} \times \vec{p}_{\ell^-} \right)  ~~, ~~ 
\overline{\tp} = \vec{p}_{\ell^{\prime +}} \cdot \left( \vec{p}_{\ell^-} \times \vec{p}_{\ell^+} \right) \label{TP1}
\end{eqnarray}
which are odd under $P$ and under $T_N$. Under $C$ and $CP$ 
they transform as: $C(\tp) =  +\overline{\tp}$, $C(\overline{\tp}) =  + \tp$, $CP(\tp) =  -\overline{\tp}$ and $CP(\overline{\tp}) =  - \tp$. 

Using the $\tp$'s in Eq.~\ref{TP1}, we define the following 
$T_N$-odd (and also $P$-violating) asymmetries: 
\begin{eqnarray}
A_T &\equiv& \frac{N\left( \tp >0 \right) - N\left( \tp < 0 \right)}{N\left( \tp >0 \right) + N\left( \tp < 0 \right)} \label{AT1} ~, \\
\bar{A}_T &\equiv& \frac{N\left( -\overline{\tp} >0 \right) - N\left( -\overline{\tp} < 0 \right)}{N\left( -\overline{\tp} >0 \right) + N\left( -\overline{\tp} < 0 \right)} \label{AT2} ~,
\end{eqnarray}
where $N\left( \tp >0 \right)$ is the number of events for which ${\rm sign}(\tp) > 0$ is measured etc. 

The asymmetries $A_T$ and $\bar{A}_T$ are sensitive to the CP-odd phase {\it BUT} are not proper CP-asymmetries, since they are not eigenstate 
of CP, i.e., $CP(A_T) = \bar{A}_T$.  In particular, we find that $A_T \propto \sin(\Delta\delta + \Delta\phi)$ and 
$\bar{A}_T \propto \sin(\Delta\delta - \Delta\phi)$, so that in general $A_T \neq 0$ and/or $\bar{A}_T \neq 0$ could be generated 
without CPV, i.e., if $\Delta\phi =0$ \& $\Delta\delta \neq 0$.
Therefore, in order to isolate the pure CPV effects, we need to combine the information from both $A_T$ and $\bar{A}_T$ as follows: 
\begin{eqnarray}
    A_{CP} &=& \frac{1}{2} \left(A_T - \bar{A}_T \right) ~.  \label{ACP1}
\end{eqnarray}

\subsection{The multi-leptons cross-section \& CP-asymmetries}\label{subsec:CPV2}

The differential (hard) cross-sections for the tri-lepton production processes under investigation, $a b \to \ell^{\prime -} \ell^+ \ell^- + X$ and 
 , can be schematically written as: 
\begin{eqnarray}
    d \hat\sigma(\ell^{\prime +} \ell^- \ell^+) = W+ U \cdot \cos(\Delta\delta + \Delta\phi) + V \cdot \tp \cdot \sin(\Delta\delta + \Delta\phi) ~,      \label{sig}
\end{eqnarray}
and similarly for the CC channel $\bar{a} \bar{b} \to \ell^{\prime +} \ell^- \ell^+ + \bar X$ with $\Delta\phi \to - \Delta\phi$ and $\tp \to \overline{\tp}$ 
in Eq.~\ref{sig}. Note that $W \propto |M_1|^2, |M_2|^2$, $U \propto {\rm Re}\left(M_1 M_2^* \right)$ and the 3rd term in Eq.~\ref{sig} (which 
is $\propto \tp$) arises 
from ${\rm Im}\left(M_1 M_2^* \right)$ and it is where the tree-level CPV resides. 
%
%

Using Eq.~\ref{sig}, we obtain for the $T_N$-odd quantities $A_T$ and $\bar{A}_T$ in Eqs.~\ref{AT1} and \ref{AT2}: 
\begin{eqnarray}
A_T = {\cal I}_{ab} \sin(\Delta\delta + \Delta\phi) ~~,~~
\bar{A}_T = {\cal I}_{\bar{a}\bar{b}} \sin(\Delta\delta -\Delta\phi) \label{ATATbar}
\end{eqnarray}
with 
\begin{eqnarray}
{\cal I}_{ab}  & \propto & \frac{\int_R d\Phi \cdot f_{a } f_{ b} \cdot V \cdot  {\tt sign}(\tp)}{\int_R d\Phi \cdot f_{a } f_{ b} \cdot \left( W+ U \cdot \cos(\Delta\delta + \Delta\phi) \right)}  ~, \label{Iab}
\end{eqnarray}
where $d \Phi$ is the phase-space volume element, $R$ is the phase-space region of integration and 
$f_a,f_b$ are the PDF's of the incoming particles $a,b$; similarly, for the CC channel, ${\cal I}_{\bar{a}\bar{b}}$ is obtained 
by replacing  $f_a f_b \to f_{\bar a}f_{\bar b}$ and $\tp \to \overline\tp$.

Thus, the CP-asymmetry $A_{CP}$ of Eq.~\ref{ACP1} is given by:
\begin{eqnarray}
A_{CP}= \frac{{\cal I}_{ab} + {\cal I}_{\bar a \bar b}}2 \cos\Delta \delta \sin \Delta \phi + \frac{{\cal I}_{ab} - {\cal I}_{\bar a \bar b}}2 \sin \Delta \delta \cos \Delta \phi ~,
\label{ACP12}
\end{eqnarray}
where the 1st term is the conventional $T_N$-odd CP-odd term which is sensitive to CPV at the tree-level (i.e., it is $\propto \cos\Delta\delta$ and therefore does not require 
a CP-even phase from FSI), while the (new) 2nd term represent a "fake" CP effect (i.e., it does not require a CP-odd phase), 
which arises when the initial state is not self-conjugate. 

To summarize this section, we have constructed three $T_N$-odd asymmetries: $A_T$, $\bar{A}_T$ and $A_{CP}$. We find that $A_{CP}$, which is supposed to be
a genuine CP-asymmetry by construction, may, also be ``contaminated'' by CP-even phases when the initial state is not CP-symmetric, 
as can be the case at the LHC.
However, Eq.~\ref{ACP12} implies that at the tree-level where $\Delta\delta =0$ (i.e., in the absence of FSI), all three asymmetries are $\propto$ CP-odd phase and are thus good measures 
of CPV, regardless of the initial state properties.

\section{The new physics framework}

We use an effective field theory (EFT) approach to describe the underlying NP responsible for CPV in tri-lepton production at the LHC 
and demonstrate our strategy for the CPV search in multi-lepton events using the following two scalar and tensor 4-Fermi effective operators:\footnote{Note that the typical  bounds on the natural scale of the operators in
 Eqs.~\ref{eq:OST} are $\Lambda \gsim {\cal O}(1)$~TeV, see \cite{ours2}.}
\begin{eqnarray}
{\cal O}_S = \left(\bar\ell_R \ell_R \right) \left( \bar t_R u_R \right) ~~,~~ {\cal O}_T = \left(\bar\ell_R \sigma_{\mu \nu} \ell_R \right) \left( \bar t_R \sigma_{\mu \nu} u_R \right) \label{eq:OST} 
\end{eqnarray}
and similarly for the $t c \ell \ell$ 4-Fermi terms, where 
\begin{eqnarray}
{\cal L}_{tu \ell\ell}^{dim6} = \frac{1}{\Lambda^2} \left( f_S {\cal O}_S + f_T {\cal O}_T \right)~, \label{Ldim6}
\end{eqnarray}
and ${\cal L} = {\cal L}_{SM} + {\cal L}_{tu \ell\ell}^{dim6}$ (we assume that the two operators have a common underlying scale). 

These 4-Fermi interactions can be generated by tree-level exchanges of heavy scalars and tensors in the underlying heavy theory.  
Interesting examples are 
the scalar leptoquarks $S_1$ and $R_2$, which transform as $(3,1,-1/3)$ and $(3,2,7/6)$, respectively, under the $\textrm{SU(3)} \times \textrm{SU}(2) \times \textrm{U}(1)$ SM gauge group.
In particular, tree-level exchanges of $S_1$ and $R_2$ among the lepton-quark pairs induce the dimension six scalar and tensor 4-Fermi operators in 
Eqs.~\ref{eq:OST}, with a specific proportion of the corresponding Wilson coefficients, obeying the relation:
\begin{eqnarray}
\vert f_T \vert = \frac{1}{4} \vert f_S \vert ~, \label{fTfS}
\end{eqnarray}
which we have used as our benchmark relation in the numerical study of the CPV effect below. 
%
%

\begin{figure}[htb]
  \centering
\includegraphics[width=0.75\textwidth]{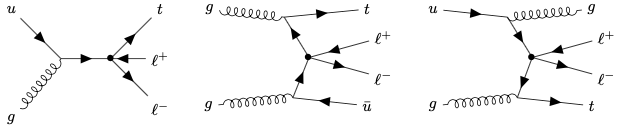}
\caption{Representative Feynman diagrams for the lowest order single top-quark + di-lepton production channels $pp \to t \ell^+ \ell^-$ and $pp \to t \ell^+ \ell^- + j$ ($j$ is a light jet), via the $t u \ell^+ \ell^-$ 4-Fermi interaction (marked by a heavy dot).}
  \label{fig:Feynman}
\end{figure}

The 4-Fermi $tu \ell \ell$ and $tc \ell \ell$ terms with complex Wilson coefficients can drive 
tree-level CPV in tri-lepton events at the LHC via the single top-quark production channels (see Fig.~\ref{fig:Feynman}):
\begin{eqnarray}
ug, gg \to \ell^+ \ell^- t,~ \ell^+ \ell^- t +j  \label{eq:tll} 
\end{eqnarray}
and the charged conjugate channel (i.e., with $\bar t$ in the final state), where $j={\rm light~jet}$.  This channel, followed by the top decay $t \to Wb \to \ell^\prime \nu_{\ell^\prime} b$, 
has interesting implications also for generic 
BSM searches of new heavy states around the TeV scale, which can generate top-leptons 4-Fermi contact terms.\cite{ours2}

\section{Collider signals of CPV in tri-lepton events}

As discussed above, we consider CPV in the single-top production channel depicted in Fig.~\ref{fig:Feynman},  
followed by the top decay to a charged lepton and focusing henceforward on the $e \mu \mu$ final state:
\begin{eqnarray}
pp \to  t \mu^+ \mu^- +X \to e^+ \mu^+ \mu^- +X ~, \nonumber \\
pp \to  \bar{t} \mu^+ \mu^- +\bar{X} \to e^- \mu^+ \mu^- + \bar{X} ~, \label{eq:emumu} 
\end{eqnarray}
where the NP and CPV phases arise from the dim.6 $tu \mu \mu$ and/or $tc \mu \mu$ 4-Fermi interactions.  

The dominant SM background to the tri-lepton production channel 
is from $WZ$ production, $pp \to WZ$, followed by the $W$ and $Z$ decays to charged leptons. 
The  SM contributions to $pp \to  e^+ \mu^+ \mu^- +X$ from other channels such as
$pp \to t \bar t, t \bar{t} V, tVV, V+jets$ ($V=Z,W$)  are much smaller, see \cite{ourpaper}. 
In particular, 
there is no interference between the $tu \mu\mu$ NP diagrams (see Fig.~\ref{fig:Feynman}) and the SM contributions to the tri-leptons signal, 
so that the cross-section is of the form: 
\begin{equation}
\sigma_{e\mu\mu}(m_{\mu\mu}^{min}) = \sigma^{SM}_{e\mu\mu}(m_{\mu\mu}^{min}) + \frac{f^2}{\Lambda^4} \sigma^{NP}_{e\mu\mu}(m_{\mu\mu}^{min})~,
\end{equation}
where we have defined the $m_{\mu\mu}^{min}$-dependent cumulative cross-section, 
selecting events with $m_{\mu\mu} > m_{\mu\mu}^{min}$: 
\begin{eqnarray}
\sigma_{e\mu\mu}(m_{\mu\mu}^{min}) \equiv \sigma_{e\mu\mu}(m_{\mu\mu} > m_{\mu\mu}^{min}) = 
\int_{m_{\mu\mu} \geq m_{\mu\mu}^{min}} d m_{\mu\mu} \frac{d \sigma_{e\mu\mu}}{dm_{\mu\mu}} ~, \label{cum} 
\label{CCSX}
\end{eqnarray}

In particular, as shown below, $m_{\mu\mu}$, which is the invariant mass of the two muons in the final state, is a useful discriminating parameter.\footnote{In the more general case of $pp \to \ell^{\prime \pm} \ell^+ \ell^- +X$, 
we have $m_{\mu\mu} \to m_{\ell \ell}$ and $m_{\ell \ell}$ would be the invariant mass of the "none-top" opposite sign same-flavor (OSSF) di-leptons from the underlying hard process, 
i.e., of the di-leptons produced from the $tt \ell \ell$ vertex and not from the top-quark decays.}

The CP-violation in our case is, therefore, a pure NP effect, since it arises from the imaginary part of the interference between the two dim.6 scalar and the tensor operators, if at least one of the corresponding Wilson coefficients  is complex.
In particular, the numerator of the CP-asymmetry $A_{CP}$ (and of the $T_N$-odd asymmetries $A_T$ and $\bar{A}_T$) is proportional to the CP-violating part of the 
cross-section for $p p \to t \mu^+ \mu^- \to e^+ \mu^+ \mu^- +X$, which is:
\begin{eqnarray}
   d\hat\sigma(CPV) \propto \epsilon \left( p_{u_i},p_{e^+},p_{\mu^+},p_{\mu^-} \right) \cdot {\rm Im} 
   \left( f_S f_T^\star \right)~, \label{sigCPV}
\end{eqnarray}
and similarly for the CC channel $p p \to \bar{t} \mu^+ \mu^- \to e^- \mu^- \mu^+ + \bar{X}$  by replacing 
$\epsilon \left( p_{u_i},p_{e^+},p_{\mu^+},p_{\mu^-} \right)$ with $\epsilon \left( p_{\bar{u}_i},p_{e^-},p_{\mu^-},p_{\mu^+} \right)$, 
where $\epsilon \left( p_1,p_1,p_3,p_4 \right) = \epsilon_{\alpha \beta \gamma \delta} p_1^\alpha p_2^\beta p_3^\gamma p_4^\delta$ and $\epsilon_{\alpha \beta \gamma \delta}$ 
is the Levi-Chivita tensor. Guided by the relation between the scalar and tensor couplings in Eq.~\ref{fTfS}, which is motivated by the matching of the EFT framework to the leptoquarks exchanges scenarios, we set below $\vert f_S \vert = 1$, $\vert f_T \vert = 0.25$ with a maximal CP-odd phase for the $t u \mu \mu$ and $t c \mu \mu$ operators, so that the CPV coupling in Eq.~\ref{sigCPV} is set to:
\begin{equation}
    {\rm Im} \left( f_S \cdot f_T^\star \right) = 0.25 ~. \label{CPVvalue}
\end{equation}

\section{Results}

All three asymmetries $A_{CP}$, $A_T$ and $\bar{A}_T$ are sensitive to the di-muons invariant mass, since 
the SM (which is dominant at low $m_{\mu \mu}$) contributes to the denominators while the CPV NP 
(which is dominant for high $m_{\mu \mu}$)  contributes to the numerators. 
We therefore exploit this $m_{\mu \mu}$ dependence to gain sensitivity to CPV. 

\begin{figure}[htb]
  \centering
\includegraphics[width=0.50\textwidth]{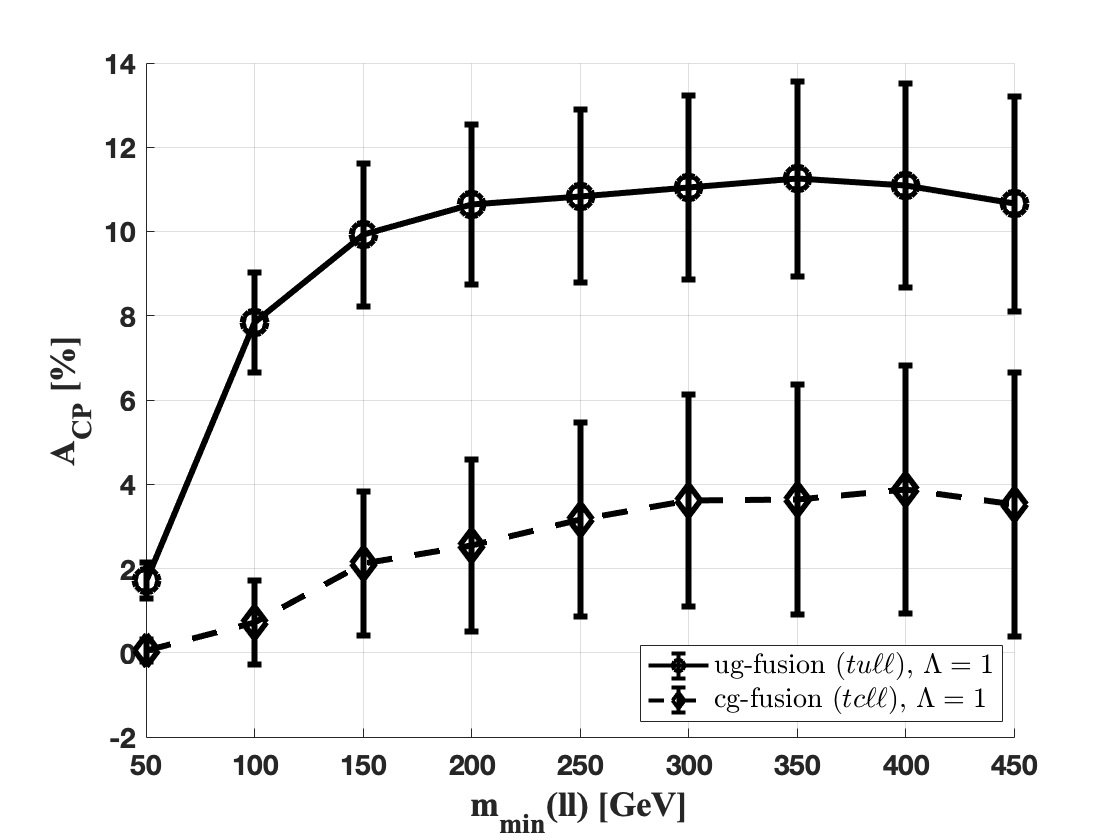}
\caption{The expected CP-asymmetry $A_{CP}$, as a function of the invariant mass cut on the same-flavor di-leptons, see text. 
The error bars represent the expected statistical uncertainty with an integrated luminosity of $1000(3000)$ fb$^{-1}$ for the ug-fusion(cg-fusion) case.}
  \label{fig:ACP}
\end{figure}

Our main results are summarized in Table~\ref{tab:data2} and Fig.~\ref{fig:ACP}:
in Fig.~\ref{fig:ACP} we show the dependence of $A_{CP}$ on $m^{min}_{\ell \ell}$ and in
Table~\ref{tab:data2} we give the resulting CP-violating and $T_N$-odd asymmetries for $m^{min}_{\ell \ell}=400$~GeV. 
Results are shown for both the $ug$-fusion and $cg$-fusion production channels, for $\Lambda=1,2$ TeV and for ${\rm Im} \left(f_S f_T^\star \right) =0.25$.
The SM background from $pp \to ZW^{\pm} +X$ is included. 
\begin{table}[htb]
\begin{center}
\caption{The expected $T_N$-odd and CP asymmetries in tri-lepton events, $pp \to \ell^{\prime \pm} \ell^+ \ell^- +X$, see text. \label{tab:data2}}
\begin{tabular}{c|c|c}
 & $ug$-fusion: $\Lambda=1(2)$~TeV & $cg$-fusion: $\Lambda=1(2)$~TeV 
\tabularnewline
\hline \hline
\
$A_{CP}$ & 11.1(7.9)\% & 3.9(0.7)\% \tabularnewline
\hline 
\
$A_{T}$ & 16.4(13.5)\% & 3.1(0.5)\% \tabularnewline
\hline 
\
$\bar{A}_T$ & -5.8(-2.3)\% & -4.7(-1.0)\% \tabularnewline
\hline \hline
\end{tabular}
\end{center}
\end{table}

\section{Summary}
Multi-leptons signals provide an excellent and rich testing ground for NP associated with flavor physics, 
lepton flavor universality, CP-violation etc...

In this work we have constructed useful CP-asymmetries for measuring
CPV effects in multi-lepton events, introducing a new modification to the classic formula for tree-level CPV in scattering and
decay processes, which takes into account the effect of an asymmetric initial state and which  
is particularly useful for CP studies at the LHC. 

Our asymmetries have several new and unique features, 
that make them particularly useful for searching for
CPV at high-energy colliders: 
\begin{itemize}
\item They use only multi-lepton final states as probes, which makes them experimentally highly distinctive.
\item They are based on simple kinematic observables that only require the reconstruction of the relatively easily-identifiable charged-lepton momenta.
\item They can be generated by tree-level CPV underlying physics, making them very sensitive to NP.
\item They are generic, since they can probe a wide range of underlying NP.
 \end{itemize}

We find that ${\cal O}(10\%)$ CP-asymmetries may be generated with new CPV TeV-scale NP, whereas  
the expected SM background for CPV in multi-lepton events is at the sub-\% level. 
Finally, with the NP considered, we expect ${\cal O}(10000)$  high-$p_T$ tri-lepton events
with an integrated luminosity of ${\cal O}(1000)$ fb$^{-1}$, if TeV-scale NP is generating the $tu\ell \ell$ 4-Fermi interactions. 
A more detailed study relevant to this talk is given in \cite{ourpaper}.

\section*{Acknowledgments}

I wish to thank the organizers of the Moriond 2023 conference for their invitation to present this talk and for 
the welcoming atmosphere in this meeting.

\end{document}